\documentstyle[prd,aps,psfig,eqsecnum]{revtex} 
\begin{document}

\twocolumn

\def\beq{\begin{equation}}
\def\eeq{\end{equation}}
\def\bea{\begin{eqnarray}}
\def\eea{\end{eqnarray}}
\def\ben{\begin{enumerate}}
\def\een{\end{enumerate}}
\def\la{\langle}
\def\ra{\rangle}
\def\a{\alpha}
\def\b{\beta}
\def\g{\gamma}
\def\d{\delta}
\def\e{\epsilon}
\def\phi{\varphi}
\def\k{\kappa}
\def\l{\lambda}
\def\m{\mu}
\def\n{\nu}
\def\o{\omega}
\def\p{\pi}
\def\r{\rho}
\def\s{\sigma}
\def\t{\tau}
\def\L{{\cal L}}
\def\S{\Sigma }
\def\gsim{\; \raisebox{-.8ex}{$\stackrel{\textstyle >}{\sim}$}\;}
\def\lsim{\; \raisebox{-.8ex}{$\stackrel{\textstyle <}{\sim}$}\;}
\def\gtrsim{\gsim}
\def\lessim{\lsim}
\def\loc{{\rm local}}
\def\vm{v_{\rm max}}
\def\bh{\bar{h}}
\def\del{\partial}
\def\nab{\nabla}
\def\half{{\textstyle{\frac{1}{2}}}}
\def\fourth{{\textstyle{\frac{1}{4}}}}

\title{Gravity with a dynamical preferred frame} 
 
\author{Ted Jacobson\thanks{jacobson@physics.umd.edu} 
and  David Mattingly\thanks{davemm@physics.umd.edu}}
\address{Department of Physics, University of Maryland,
College Park, MD 20742-4111, USA}
\maketitle
\begin{abstract}
We study a generally covariant model in which 
local Lorentz invariance is broken   
by a dynamical unit timelike vector field $u^a$---the ``aether".
Such a model makes it possible to study the gravitational 
and cosmological consequences of preferred frame effects,
such as ``variable speed of light" or high 
frequency dispersion,
while preserving a generally covariant metric theory of gravity.
In this paper we restrict attention to an action for 
an effective theory of the 
aether which involves only the antisymmetrized derivative
$\nab_{[a}u_{b]}$. Without matter this theory is equivalent
to a sector of the Einstein-Maxwell-charged dust system.
The aether has two massless transverse excitations, and 
the solutions of the model include all vacuum solutions of
general relativity (as well as other solutions). However, the
aether generally develops gradient 
singularities which signal a breakdown of this effective theory. 
Including the symmetrized derivative 
in the action for the aether field may cure this problem.

\end{abstract}
\section{Introduction}

The Lorentz group is non-compact, since the boost parameter 
is unbounded. This makes exact Lorentz invariance impossible 
to test uniformly. 
Lorentz invariance has thus been tested only up to some maximum
boost and beyond that lies an infinite volume of uncharted territory
in the Lorentz group.
Contrast this with the rotation
group. Rotation invariance can be tested by filling in the compact 
$SO(3)$ group manifold more and more densely with data points,
or by checking a few randomly selected rotations. The rotation
group can be and has been uniformly explored.

There is also reason to doubt exact Lorentz invariance: 
it leads to divergences in quantum field theory associated 
with states of arbitrarily high energy and momentum.   
This problem can be cured with a short distance cutoff
which, however, breaks Lorentz invariance.

For these reasons we entertain the possibility 
that there is a preferred rest frame at each spacetime point.
In particular, we seek a viable effective field theory
incorporating a breaking of local Lorentz invariance.

If the preferred frame were to be a fixed external structure,
then it would violate general covariance, which would  
require us to abandon general relativity (or
any generally covariant modification thereof).
General covariance ordinarily implies that the divergence
of the matter energy-momentum tensor $T_{ab}$ vanishes when
the matter field satisfies its equation of motion.
This is required for consistency of the Einstein
field equation\footnote{We use units with $c=1$ and the
metric signature $({+}{-}{-}{-})$.}
$G_{ab}=8\pi GT_{ab}$, since the 
divergence of the Einstein tensor $G_{ab}$ is identically zero
by virtue of the contracted Bianchi identity.
If a fixed preferred frame is introduced into the
matter action, for example, general covariance is lost, 
$T_{ab}$ is not divergenceless, and the Einstein
equation is inconsistent. 
We therefore seek to incorporate
the preferred frame while preserving
general covariance, which requires that the preferred frame
be {\it dynamical}. 

It would be most appealing if the preferred frame were 
somehow determined by the spacetime metric itself. 
As discussed below, 
in spacetimes with an initial singularity
the metric can be used to define a cosmological 
time function, the gradient of which then determines
a preferred frame (which is by construction timelike).
However, the nonlocal relationship between
this frame and the metric results in infinitely nonlocal
field equations if this frame is incorporated into the action
principle. Theories with even a finite amount of nonlocality
are generally horribly unstable\cite{woodard}, so we do not 
consider this a viable approach.

To avoid such unacceptable behavior the preferred frame should
arise from {\it local conditions}, which of course 
reflect conditions at
earlier times but only through {\it dynamics}.
For example, this dynamical frame could 
be defined by a vector field\cite{willnord,nordwill,hellnord,moffatv}
or by the gradient of a scalar field\cite{moffats,vslc}.
In these approaches the presence or absence
of a preferred frame depends on the field configuration,
since the preferred vector may vanish or may not be timelike.

Since our motivation arises from doubts about the 
fundamental validity of exact Lorentz invariance,
we are not interested in a theory possessing a Lorentz invariant
phase. We wish to study instead an effective 
theory in which there is {\it always} a 
preferred frame. 
This frame is defined by a timelike direction
or, equivalently, by a unit timelike 
(contravariant or covariant)  
vector field.\footnote{A unit 
timelike vector contains a discrete 
piece of information that a frame by itself does not have,
namely, a time orientation. The theory may or may not 
depend on this orientation.}
Such a field carries a nonlinear representation of the local 
Lorentz group since the field takes values not in a vector
space but on the unit hyperboloid in the tangent space.
This could therefore be called a theory of nonlinearly realized 
Lorentz invariance. It is analogous to a gauge theory
with a non-linear sigma model Higgs field of fixed 
norm\cite{quaternion,appelquist}. 
There seems to be no
generally accepted terminology for this sort of symmetry
structure. 
Since the symmetry breaking unit vector field is not
a state-dependent expectation value but rather   
breaks this symmetry in {\it all} states, it may be misleading to 
say the symmetry is ``spontaneously broken". 
For lack of a better
idea we shall take refuge in ambiguity and just call it 
"broken". What this really means is that in order to 
implement the local Lorentz symmetry one must transform not only
the matter fields but also the background unit vector, so for 
practical purposes it will appear as what would normally
be called broken Lorentz symmetry. 

The theory described here was devised about a decade
ago by J.C. Dell together with one of us\cite{Dell},
and we have since learned that similar ideas
have independently been studied previously.
The non-gravitational part of the 
theory (and generalizations thereof) was considered by 
Dirac\cite{Dirac} in the early 1950's as a new theory of electrons
(in which the unit timelike vector played the dual role of 
gauge-fixed vector potential and flow vector of a stream of 
charged dust).  A class of generally covariant theories
breaking Lorentz invariance was 
studied by Gasperini\cite{Gasperini} in many papers.
In this work the tetrad formalism was used, and the local
Lorentz symmetry was broken by including in the action terms referring to 
a fixed ``internal" unit timelike vector. This is equivalent to
our formulation in terms of the metric and a unit timelike vector.
To see the equivalence, note that the tetrad defines a metric and
associates to the fixed internal vector a unit vector field on spacetime. 
The only other information in the tetrad is the gauge freedom parametrized 
by the local rotations leaving invariant the preferred 
timelike vector. Eliminating this gauge freedom leads to the formalism
used in this paper.  Gasperini has studied both cosmological and 
central field solutions, with various choices for 
the specific form of the second derivative terms in the Lagrangian. 
In the present paper we focus on a different Lagrangian.

The particular Lagrangian studied here  was also considered 
by Kosteleck\'y and Samuel\cite{KosSam},
as a simplified model of the spontaneous Lorentz symmetry
breaking that might occur in string theory, although in 
Ref. \cite{KosSam} the preferred vector was 
not necessarily timelike.  
More generally, those authors argued that spontaneous
Lorentz symmetry breaking in string theory may produce
vacuum expectation values of more than one tensor field.
In this case, rather than having just a single
``preferred frame" there might be several background Lorentz
tensors which collectively break part or all of the Lorentz
symmetry.

The remainder of this paper is organized as follows.
In section \ref{cosmo} we explain the nonlocality problem encountered
if a cosmological time is used to define the preferred frame.
In section \ref{aether} our proposed field theory of a preferred frame
is formulated and its general properties are investigated.
It is seen that the solutions for our theory comprise a subset
of the solutions to the coupled Einstein-Maxwell-charged dust equations.
Several types of exact solutions to the field equations are 
characterized in section \ref{solutions}, and the linearized theory
is studied in section \ref{linearized}. Coupling of the preferred 
frame to matter fields is discussed in section \ref{matter}.
Both the dimension $\le4$ couplings
and some higher dimension ones are examined.  
The paper concludes with a brief discussion in section \ref{discussion}.

Among the dimension $>4$ couplings are included 
theories involving Lorentz non-invariant dispersion at 
high wavevectors. 
This is motivated by recent work in which high frequency
dispersion is invoked to avoid the role of trans-Planckian
modes in the Hawking effect (for a review see \cite{river}).
In this framework such theories can be formulated
in a generally covariant manner so that gravitational 
effects can be consistently incorporated. 
Higher dimension couplings 
also also provide an alternative generally
covariant formulation of variable speed of light models,
in which different fields propagate at different speeds
possibly at different cosmological epochs.
Such models have recently been of interest as potential
alternatives to standard cosmology, and have been
given generally covariant formulations using additional
fields to define the preferred frame\cite{moffatv,moffats,vslc}.

\section{Cosmological times}
\label{cosmo}

In this section we briefly describe the construction
of a cosmological time function determined purely by 
the metric, and the reason for rejecting
it for the purposes of an effective theory of local Lorentz
symmetry breaking. 

The cosmological metric of our universe, by
virtue of its (approximate) homogeneity, 
(approximately) defines a preferred
spacelike foliation of spacetime.
However, this particular definition of the 
time function relies on the symmetry of the 
spacetime. For a workable theory with general
covariance what is needed is a definition of 
cosmological time that can be used independently of 
symmetry. 

It is difficult to think of a notion of cosmological
time that would make sense for all possible cosmologies.
However, if we restrict attention to spacetimes with
a ``beginning",
then two notions of cosmological 
time at $x$ present themselves: ($i$) {\it volume time},
the spacetime volume 
(or perhaps the fourth root thereof) of the past set $I^-[x]$, 
and ($ii$) {\it maximal time}, 
the maximal proper time along a causal 
curve going back to the initial 
singularity.\footnote{The
maximal time function has been discussed in Refs. \cite{BT,WY,AGH}.
In particular, a powerful theorem proved in 
Ref.\cite{AGH} establishes a number of properties
of this function under the further 
assumption that the initial singularity is the only place 
past directed causal curves can end.}
Other possibilities are combinations or smoothed averages
of these times.
Both of these time functions are determined non-locally
but causally by the spacetime to the past of $x$.
They may or may not be sufficiently smooth functions 
to enter meaningfully into a local action 
principle.\footnote{One of the results of the
theorem of Ref. \cite{AGH} referred to in the previous 
footnote 
is that the maximal time function is locally Lipschitz
and its first and second derivatives exist almost everywhere.
The volume time function may well be even better behaved.}
If we assume that they are indeed sufficiently smooth
we find that there is in any case a fatal problem with using
them in this manner, as will now be explained.

Suppose that to the usual action for gravity and matter 
fields is added a term involving one of the above 
cosmological times, 
\beq
S= S_{\rm local} + S_\t.
\eeq
We assume that the equations of motion are obtained
as usual by requiring that the action is stationary 
with respect to variations of the fields. 
The variational derivative of the action with respect to 
$g_{ab}(x)$ has the form
\beq
\frac{\d S}{\d g_{ab}(x)} =  
\frac{\d S_\loc}{\d g_{ab}(x)} 
+\frac{\d' S_\t}{\d g_{ab}(x)}
+\int d^4x' \frac{\d S_\t}{\d \t(x')}
\frac{\d\t(x')}{\d g_{ab}(x)},
\eeq
where the prime on $\d'$ indicates that the metric dependence
of $\t(x)$ is {\it not} included in the variation.
Since $\d\t(x')/\d g_{ab}(x)$ has support when 
$x'$ lies to the future of $x$, the field equation 
${\d S}/{\d g_{ab}(x)}=0$  
involves the values of the fields to the future of $x$. 
Indeed the metric field equation is infinitely non-local
in time, since the time function at any point to the future
can be affected by a metric variation at $x$.
Even finite nonlocality in time leads to unphysical instability
\cite{woodard}, so this approach to incorporating a preferred
frame must be rejected.
If the action depends on $\t$ only through its
derivative $\nabla_a\t$, then
the equation of motion would be causal if 
$\d\nabla_c\t(x')$ depended only on $g_{ab}(x)$ at $x=x'$.
However, this is not the case for the either the
volume time or the maximal time. 

\section{Aether dynamics}
\label{aether}

We now turn to a class of theories in which 
there is a preferred frame which is 
determined by a local field.
It is convenient to give a name to this field, 
and ``aether" seems as good a name as any. 
Let us take the aether field to be a 
unit timelike vector field $u^a$, which 
is dimensionless, like the metric.
To handle the condition that $u^a$ is a {\it unit}
vector, we include in the action a Lagrange multiplier term.
Note that we are implicitly assuming that the spacetime 
admits a globally defined unit timelike vector field
which is the case if and only if the spacetime is time orientable.

\subsection{Action}
The most general Lagrangian involving the metric
and the aether with two or fewer derivatives is,
up to a total divergence,
\bea
{\cal L}_{g,u} &=& 
a_0 
- a_1 R
- a_2 R_{ab}u^au^b\nonumber\\
&&
- b_1\, F^{ab}F_{ab} 
- b_2\, (\nabla_a u_b) (\nabla^a u^b)
-b_3\, \dot{u}^a\dot{u}_a,
\label{Lgu}
\eea
where $\dot{u}^a:= u^m\nabla_m u^a$, and 
$F_{ab}$ is defined in analogy to the 
electromagnetic field strength,
\beq
F_{ab}:= 2\nabla_{[a} u_{b]}.
\eeq 
The term $(\nabla_a u^a)^2$ is equivalent,
via integration by parts, to the combination
$(\nabla_a u_b) (\nabla^a u^b)-(1/2)F^{ab}F_{ab} 
+ R_{ab}u^au^b$, so has not been included in (\ref{Lgu}).

The Lagrangian (\ref{Lgu}) is similar to the one discussed in
Ref.   \cite{Will} as the most general Lagrangian for a
vector-tensor theory of gravity including terms up to
second order in derivatives and   quadratic in the vector
field. The differences are that (i) the terms $u^2$ and
$Ru^2$ are missing from our action since the vector field is
constrained to be a unit vector, and (ii) we have included
the quartic term $b_3 \dot{u}^2$, which was omitted in Ref.
\cite{Will} because it is not quadratic in
$u^a$.\cite{delludot} Note that, even without the last
term, our theory is {\it not} a special case of the
vector-tensor theories discussed in Ref. \cite{Will}, since
the constraint $u^2=1$ affects the field equations.

The coefficient $a_0$ in the action (\ref{Lgu}) 
has mass dimension 4 while $a_{1,2}$ and $b_{1,2,3}$
have mass dimension 2. Lacking the underlying 
fundamental theory we do not try to assign {\it a priori} 
the values of these coefficients. A partial analysis of 
the observational consequences and limits on them
has been done for the vector-tensor theories\cite{Will},
however that analysis does not apply directly to our
case due to the presence of the constraint term.
It is fairly clear nevertheless that whatever values
$b_{1,2,3}$ take, agreement with observation
will require that $a_2\ll a_1$,
and that $a_0/a_1$ (which is basically the cosmological constant)
must not be much larger than the squared Hubble constant. 

In this initial foray we shall restrict attention 
to the simple case in which the only terms with non-zero
coefficients are $R$ and $F^2$. 
That is, we set $a_0=a_2=b_2=b_3=0$.
The minimal theory we consider is thus 
defined by the action
\bea
S_{min}[&&g_{ab},u^a,\l]=
\int d^4x\, \sqrt{-g}\, \nonumber\\
&&~~\Bigl(
-a_1 R 
- b_1\, F^{ab}F_{ab}
+\l(g_{ab}u^a u^b -1)
\Bigr).
\label{Smin}
\eea
This minimal theory is one of the models considered by
Kosteleck\'y and Samuel\cite{KosSam} in the paper mentioned
in section I. Those authors studied a broader class
of models in which $u^a$ is not necessarily constrained to be a 
unit vector but rather possesses a Lorentz-invariant potential 
energy with a minimum at some fixed value of $u_au^a$. They
also allowed for extra, compact spatial dimensions of spacetime, and
examined cases where the symmetry breaking vector lies in 
the extra dimensions as well as cases where it lies in the 
four ordinary spacetime dimensions. Our paper by contrast
is restricted to four dimensions and to a timelike vector of 
fixed norm.
Later in this paper we shall also add matter terms to the action, 
including terms which couple the aether field to the matter.

Note that $F_{ab}$ is invariant under the 
``gauge transformation" 
\beq
u_a\rightarrow u_a + \nabla_a f,
\label{gauge}
\eeq
however the constraint $ u^2=1$
does not share this symmetry
(nor do the additional couplings 
in general), so the theory is certainly not 
``gauge invariant". 
The constraint does have a limited version
of this symmetry however, 
namely for those functions $ f$ satisfying
$(u_a+\nabla_a f)(u^a +\nabla^a f)=u_au^a=1$.
The general solution to this equation is 
\beq
u^a\nabla_a f=-1\pm\sqrt{1+q^{ab}\nabla_a f\nabla_b f},
\label{fdot}
\eeq
where 
\beq
q_{ab}:=-g_{ab}+u_au_b
\label{q}
\eeq
is the (positive definite) spatial metric orthogonal to $u^a$.
Thus the action (\ref{Smin}) is invariant 
under the gauge transformation (\ref{gauge})
if $ f$ is chosen arbitrarily on a spacelike
surface and then determined uniquely elsewhere
(up to a discrete choice of sign) 
by integration of (\ref{fdot}) 
along the flow of $u^a$.

\subsection{Field equations}
The equations of motion arising from the action
(\ref{Smin}) are
\beq
G_{ab}= -\frac{2b_1}{a_1}
(F_{am}F_b{}^m-\fourth F^2 g_{ab}) 
+\frac{\l}{a_1}u_au_b,
\label{geq}
\eeq
\beq
\nabla_a F^{ab} = -\frac{\l}{2b_1}\, u^b,
\label{ueq}
\eeq
\beq
g_{ab}u^a u^b =1.
\label{leq}
\eeq
The metric equation (\ref{geq}) has the form 
of the Einstein equation
$G_{ab}=8\pi G\, T_{ab}$, where $G=1/16\pi a_1$,
and the stress tensor receives contributions 
from both the $F^2$ term and the constraint term in the action.
(The constraint equation (\ref{leq}) has been used 
to drop the contribution to  
(\ref{geq}) that would have come from the variation
of $\sqrt{-g}$ in the constraint term.)
The contribution from the constraint term
looks like that of a (pressureless) dust with rest energy density
$2\lambda$, and that from the $F^2$ term 
is the usual Maxwell tensor familiar from electromagnetism, 
if we identify the vector potential as 
\beq
\mbox{``}A_m\mbox{"}\leftrightarrow 2\sqrt{b_1}u_m. 
\label{A}
\eeq
The stress tensor thus satisfies the usual energy conditions
provided $b_1/a_1$ and $\l/a_1$ are positive.

In terms of the vector potential $A_m$ (\ref{A})
the constraint equation (\ref{leq}) becomes 
\beq
A_m A^m = 4b_1
\label{Aconstraint}
\eeq
which can be interpreted as a gauge condition.
The aether field equation (\ref{ueq})
becomes the Maxwell equation with source equal to the 
current of a charged dust fluid with 4-velocity $u_b$ and 
charge density $-(\l/\sqrt{b_1})$. The evolution of $\l$
is determined by the current conservation equation which
follows from divergence of the aether field equation (\ref{ueq})
upon using the identity $\nabla_a\nabla_bF^{ab}\equiv0$. 
Thus $\l$ satisfies
a first order ordinary differential equation along the 
flow lines of $u^a$:
\beq
u^a\nabla_a\l=-\l\, \nabla_a u^a.
\label{ldot}
\eeq
In particular, if $\l$ vanishes on a
Cauchy surface, it must vanish everywhere. Also, the sign
of $\l$ on a given flow line cannot change, since if $\l=0$
at any point on a flow line it must vanish everywhere
on that line.

\subsubsection*{Relation to Einstein-Maxwell-charged dust system}
We have just seen that 
the field equations of the minimal theory 
take the form of the coupled Einstein-Maxwell equations, with
a charged dust matter source possessing charge to mass
ratio $-1/2\sqrt{b_1}$. There is no explicit 
equation of motion for the dust, however the normalization condition 
(\ref{leq}) provides such an equation. Taking the gradient of  
$u^2=1$ we have  
\bea
0&=&\nabla_a (u^bu_b)\\
&=& 
2u^b\nabla_a u_b\\ 
&=& 2(u^b\nabla_b u_a +u^bF_{ab}).\label{dusteq}
\eea
Let us define $\tilde{F}_{ab}=2\del_{[a}A_{b]}=2\sqrt{b_1}F_{ab}$.
Then (\ref{dusteq}) becomes
\beq
u^b\nabla_b u_a = -\frac{1}{2\sqrt{b_1}}\tilde{F}_{ab} u^b,
\label{Lorentzforce}
\eeq
which is the equation of motion for a particle in the electromagnetic
field $\tilde{F}_{ab}$, with charge to mass ratio $-1/2\sqrt{b_1}$,
the same ratio we inferred from the Einstein and Maxwell equations!
Thus any solution of our minimal theory is a solution of the 
Einstein-Maxwell-charged dust equations (although the converse is
not true). 
The equivalence to a subset of the charged dust solutions demonstrates
that the equations of our theory admit an initial value formulation,
and it provides some useful intuition about the nature of the solutions.

Our theory is {\it not equivalent} to the 
Einstein-Maxwell-charged dust system because in the general
solution of that system the dust 4-velocity is not 
proportional to the vector potential in some gauge.
That is, although there is always a gauge transformation that 
will make $A_m/2\sqrt{b_1}$ a unit vector, it cannot in general 
be made to coincide with the dust 4-velocity.\footnote{The 
general form of the discrepancy between these two 4-vectors was found 
by Dirac (see the second paper of Ref.\cite{Dirac}), who showed that
(in four spacetime dimensions) there is always a gauge in which 
$A_m/2\sqrt{b_1}= u_m + \xi\nab_m\eta$, 
where $\xi$ and $\eta$ are functions that are constant along the
flow lines of $u^a$. Dirac included the functions $\xi$ and $\eta$ 
as dynamical variables in
order to obtain a theory in which arbitrary electron
streams were admitted. In the third paper of Ref.\cite{Dirac} he 
allowed for multiple streams.}

\section{Solutions}
\label{solutions}
In this section we characterize a few types of solutions
to the field equations.

\subsection{Solutions with $\l=0$}
\label{l=0}
If $\l=0$, then the two field equations
(\ref{geq},\ref{ueq}) are just the Einstein-Maxwell 
equations. Any solution to these equations is
a solution in our theory provided a gauge can
be chosen so that the constraint equation (\ref{Aconstraint})
is satisfied. Such a gauge always exists,
at least locally.

\subsection{Solutions with $F_{ab}=0$}
\label{F=0}
A special class of solutions to the field equations 
with $\l=0$ are those with $F_{ab}=0$.
For such fields, (\ref{ueq}) implies 
that $\l=0$, and the 
field equations (\ref{geq}-\ref{leq}) reduce
to the ordinary vacuum Einstein equation together 
with the constraint $u^2=1$. 
When $F_{ab}=0$ it follows, at least locally, that 
$u_a=\nabla_a\t$ for some function $\t$, and the constraint
then implies that $\nabla_a\t \nabla^a\t=1$. The general
solution for such a function $\t$ can specified by 
assigning the value $\t=0$ to an arbitrary spacelike surface,
and determining $\t$ elsewhere by ``uniform normal extension",
i.e. by the differential equation 
$n^a\nabla_a\t=1$, where $n^a$ is the unit normal to the
surface. 

Another way to think of this construction is in terms
of the congruence of integral curves of $u^a$. 
When $F_{ab}=0$, eqn. (\ref{dusteq}) implies that these
curves are geodesics. 
Moreover, if $u^a$ is the unit tangent field to a congruence
of geodesics, then $F_{ab}= 2\nab_{[a} u_{b]}=0$ 
if and only if the congruence
is hypersurface-orthogonal. Hence the general solution
of this type is just an arbitrary solution to the 
Einstein equation, together with $u^a$ given by 
the unit tangent field
of any hypersurface-orthogonal congruence of timelike geodesics
in this metric. A special case is flat spacetime, where
the $u^a$ congruence consists of straight lines normal to 
an arbitrary initial spacelike hypersurface. 

\subsubsection*{Singular aether evolution}
This characterization of the $F_{ab}=0$ solutions 
shows that, at least for such solutions,
the evolution of $u_a$ is generally singular.
The geodesics launched normally from a 
spacelike surface will typically cross. Where they
do, the quantity $\nabla_a u_b$ will diverge. 

The existence of such singular evolutions for $u_a$
signals a breakdown of the effective theory
we are using. Perhaps it would be cured by 
including the term $(\nabla_{(a}u_{b)})(\nabla^{(a} u^{b)})$ 
in the action. (Without this term the action is insensitive to 
gradients for which the antisymmetrized derivative
$\nab_{[a} u_{b]}$ vanishes.)
For the purposes of the 
present paper we shall not pursue this question, 
but it should be addressed. 

\subsubsection*{Cosmological solutions}
If $u^a$ shares the symmetry of 
a homogeneous isotropic cosmological metric, then
$F_{ab}=0$. The presence of the aether field therefore 
has no influence on the cosmological evolution
unless there are additional terms in the 
action beyond the minimal model.
In Ref. \cite{nextpaper} 
we examine some cosmological effects of 
coupling to a scalar field though a fourth spatial
derivative term as discussed in Sect. \ref{disp}. 

\subsubsection*{Black hole solutions}
For a spherically symmetric black hole, 
a suitable congruence of geodesics is given by the
radial free-fall trajectories that all have the same 
Killing energy, i.e. the same asymptotic velocity at
spatial infinity. The same construction can even 
be applied in the case of a Kerr black hole, at least
for the geodesics that are at rest at spatial infinity.
This follows from the work of Ref. \cite{Doran}, in
which this congruence is employed to construct
a coordinate system for the Kerr metric using the 
time function $\t$ mentioned above.

\subsection{Spherically symmetric, static solutions}
Here we seek to characterize the general
spherically symmetric, static solution.
We shall find that, besides the mass, the metric
in these solutions has an additional free parameter, 
the ``aether charge".

Some of the linearized static, spherically
symmetric solutions were previously studied in Ref. \cite{KosSam}.
Those authors examined the case where $u^a$ is spacelike, and 
while in four spacetime dimensions they restricted
attention to vanishing Lagrange multiplier
$\l$ and vanishing field strength $F_{ab}$. 
In the present work we treat the nonlinear case,
considering only timelike $u^a$
and imposing no further restrictions on the fields. 

Coordinates can be chosen so the line element
takes the form 
\beq
ds^2 = g_{tt}dt^2 + g_{rr}dr^2-r^2(d\theta^2
+ \sin^2\theta\, d\phi^2),
\label{gspherical}
\eeq
and the aether field takes the form 
\beq
u = u_t(r) dt + u_r(r)dr.
\label{uspherical}
\eeq
The only potentially nonzero component of $F_{ab}$
is then $F_{rt}=\partial_r u_t$, and 
the constraint (\ref{leq}) implies
\beq
g^{tt}u_t^2 + g^{rr}u_r^2 = 1.
\label{cspherical}
\eeq

The aether field equation (\ref{ueq}) in coordinate
form reads
\beq
\frac{1}{\sqrt{-g}}\partial_\a
\left(\sqrt{-g}g^{\a\m}g^{\b\n}F_{\m\n}\right)
=-\frac{\l}{2b_1}u^\b,
\label{ueqcoord}
\eeq
or, taking into account the form of the metric
(\ref{gspherical}),
\beq
\frac{1}{\sqrt{-g}}\partial_r
\left(\sqrt{-g}g^{rr}g^{\b t}F_{rt}\right)
=-\frac{\l}{b_1}u^\b.
\label{ueqspherical}
\eeq
The left hand side vanishes when $\b=r$, hence
the field equation implies that
$\l u_r=0$, which in turn implies that either
$u_r=0$ or $\l=0$. In the former case, 
$u^a$ is proportional to the timelike Killing field
itself.
There are thus two cases to consider.

If $\l\ne0$ there are in fact no static solutions, unless the 
coefficients in the action are such that the charge
to mass ratio of the dust is extremal. 
Recall that any solution to our theory is a solution
to the charged dust theory. However under
the influence of gravitational and electric forces, the
non-extremal charged dust cannot remain static, since there is 
no pressure\cite{Bonner,DeRay}. 

If $\l=0$ then (cf. Sect. \ref{l=0})
these are just the spherically
symmetric static Einstein-Maxwell 
solutions, i.e. the Reissner-Nordstrom
solutions, in a spherically symmetric,
static gauge with fixed norm (\ref{Aconstraint}).
Such a gauge always exists, at least locally.
Consider the gauge transform 
$A_\m=A'_\m+\a_{,\m}$ of an arbitrary vector potential
$A'_\m$. To maintain spherical symmetry and time
independence we must have $\a=\b t + \g(r)$ 
(using the coordinates in (\ref{gspherical})), so that
$A_t=A'_t + \b$ and $A_r =A'_r +\g_{,r}$.
The normalization (\ref{Aconstraint})
then implies
\beq
\g_{,r} = -A'_r \pm 
\sqrt{g^{tt}\bigl[g^{tt}(A'_t+\b)^2-4b_1\bigr]},
\eeq
where we have used the fact that for the 
Reissner-Nordstrom metrics $g_{rr}=-1/g_{tt}$.
In any region where $g^{tt}$ and $A_t$ are bounded
one can always choose $\b$ large enough so that
the radical is real, and then $\g(r)$ can be found
by integration. 

If a horizon is present $g^{tt}$ diverges and 
it is not clear from the preceding
discussion whether the unit timelike gauge can be 
accessed in a smooth way across the horizon. 
Indeed it can, however the maximal extension of the 
region over which such a gauge can be accessed 
depends on the parameter $b_1$. This can also be understood
from the equivalence with a charged dust solution.
The radial congruence $u^a$ must satisfy the Lorentz force equation 
(\ref{Lorentzforce}), and this congruence can be 
nonsingular and time-independent only if the trajectories
are monotonic in the $r$ coordinate.
In general, however, the trajectories bounce inside the black hole.

\subsubsection*{Comparison with observation}
To compare with observation it would be necessary to determine 
which of the above solutions to use in the presence of a 
spherically symmetric static source such as a planet, star,
or black hole. The metric associated with one of 
these objects depends on its ``aether charge"
---the charge of the Reissner-Nordstrom solution---which 
is determined by the ``charge" of the ``aether dust" that fell in
when the object condensed.
The choice is determined by the initial conditions on $\l$, 
which are presumably cosmological in origin. We have no
theory of these initial conditions at this stage, but 
agreement with observations can put a bound on the amount
of aether charge. 
If this charge is zero, then we have the usual Schwarzschild
solution of general relativity (and, as discussed in section
\ref{F=0}, the aether field is the tangent field to a hypersurface
orthogonal congruence of timelike geodesics), 
which of course agrees with observations. 

\section{Linearized theory}
\label{linearized}

In this section we study the linearized equations
defined by expanding about a background solution,
\begin{eqnarray}
g_{ab}&=& g^{(0)}_{ab} +h_{ab}\\
u_a &=& u^{(0)}_a + v_a,\\
\l &=& \l^{(0)} + \l^{(1)}.
\end{eqnarray}
For the background we take the flat metric
$g^{(0)}_{ab}=\eta_{ab}$ and a constant $u^{(0)}_a$. 
In this background solution
the equations of motion imply that the Lagrange multiplier 
$\l^{(0)}$ must vanish, hence we shall use the letter $\l$ for
the perturbation $\l^{(1)}$. In this section we use the
flat background metric to raise and lower indices.
Note that we use the perturbation of the {\it covariant}
vector $u_a$ to define the perturbation $v_a$.

We choose cartesian coordinates $(x^0,x^i)$, $i=1,2,3$, in which
the components of $\eta_{ab}$ are 
$\mbox{diag}(1,-1,-1,-1)$ and those of $u^{(0)}_a$ are 
$(1,0,0,0)$. 
The linearized field equations for
this theory were first written down in a general gauge 
in Ref.\cite{KosSam}, who also pointed out that the Lorentz 
gauge can be accessed using the linearized
diffeomorphism invariance of the action.
The Lorentz gauge condition for the metric perturbation is
$\bh_{ab}{}^{,b}=0$, where $\bh_{ab}\equiv h_{ab}-\half h \eta_{ab}$
is the trace-reversed metric perturbation. 
In this gauge, the linearized equations of
motion are
\beq
\Box \bh_{ab}  
=2  \frac {\l} {a_1} u^{(0)}_a u^{(0)}_b\label{heq}
\eeq
\beq
\Box v_b -\del_b (\del^a v_a)= -\frac{\l}{2 b_1} u^{(0)}_{b}\label{veq}
\eeq
\beq
-h_{ab} u^{(0)a} u^{(0)b} + 2 u^{(0)b} v_b = 0\label{linleq}
\eeq

In a source-free region, the residual gauge freedom 
is usually employed to set $\bh_{0i}=h_{0i}=0$ and 
$\bh=-h=0$.\cite{wald}
The possibility of doing so depends on the fact that these
quantities satisfy the wave equation in general relativity. 
In our case, the components $h_{0i}$ satisfy the wave equation, 
however $h$ does not, due to the source term on the right
hand side of (\ref{heq}) that is always present 
(even outside matter) unless $\l=0$. This source corresponds to the
energy density of the ``charged dust", and we wish to allow for 
the presence of this term.
Therefore, rather than setting $h$ to zero, we choose to set
to zero the trace of the spatial part,
$\bh^i_i$, which does satisfy the wave equation.
The proof that this can be done follows the same logic as 
in the usual case.
This gauge condition  implies $\bh=\bh_{00}$, hence
$h_{00}=\bh_{00}-\half\bh\eta_{00}=\half\bh_{00}$,
and the Lorentz gauge condition 
implies $\del_0 \bh_{00}=0$ and $\del_i \bh_{ij}=0$.
Thus $h_{00}$ is time-independent and
the spatial part $\bh_{ij}$ is a transverse traceless
solution to the wave equation. This is not quite the same
as the usual transverse traceless gauge in general relativity
however, since $h_{ij}=\bh_{ij}-h_{00}\d_{ij}$, so $h_{ij}$ is 
not transverse unless $\del_i h_{00}=0$, and 
$h=-2h_{00}\ne0$.

It remains to consider the linearized equations for 
$h_{00}$, $v_a$, and $\l$. The 
$00$-component of the metric equation (\ref{heq})
and the constraint equation (\ref{linleq}) 
determine $\l$ and $v_0$ in terms of $h_{00}$:
\bea
\l &=& -a_1\nab^2 h_{00}\\
v_0 &=&-\half h_{00}.
\eea 
Using the time-independence of $v_0$, 
the time and space components of the aether equation 
(\ref{veq}) read
\bea
-\nab^2 v_0 -\del_0(\del^i v_i) &=& -\frac{\l}{2 b_1}\label{v0eq}\\
\Box v_i -\del_i (\del^j v_j)&=&0.\label{vieq}
\eea
Let us decompose $v_i$ into transverse and longitudinal parts,
$v_i = v^T_i + v^L_i$, where 
$\del^iv^T_i=0$ and $v^L_i = \del_i f$ for some scalar
field $f$. Then (\ref{vieq}) implies that the transverse
part satisfies the wave equation, $\Box v^T_i=0$,
so the the aether field has two transverse massless modes.

As for the longitudinal part $v^L_i$,
equation (\ref{v0eq}) implies
\beq
\del_0(\del^i v^L_i)=
\frac{b_1-a_1}{2b_1}\nab^2 h_{00},
\eeq
so in particular $v^L_i$ has at most linear time dependence.
(The same conclusion follows from the divergence of 
Eq. (\ref{vieq}).)
Thus $v^L_i = \del_i a(x)\, t + \del_ib(x)$,
where $a(x)$ is determined by $h_{00}$ and $b(x)$ is arbitrary. 

In summary, the perturbation spectrum consists of two
massless transverse traceless modes of $\bh_{ij}$,
and two massless transverse modes of $v_i$. In addition
there is a mode in which $h_{00}$ is an 
{\it arbitrarily} specified
time-independent function, which determines $v_0$,
$\l$, and the time derivative of the longitudinal part
$v^L_i$. The time-independent part of $v^L_i$ is also arbitrary.
This last freedom corresponds to the linearization of the 
restricted gauge symmetry (\ref{fdot}). 

The longitudinal mode looks very strange at first sight.
In the charged dust interpretation, the dust energy density
is adjusted to produce an arbitrary gravitational potential
$h_{00}$, and the perturbed metric, electromagnetic field,
and charge density are 
all time independent, while the perturbed dust world lines
are time dependent. 
This is a peculiarity of the first order perturbative solution
however. No exact solution shares this
property, as can be easily seen from the aether 
field equation (\ref{ueq}).
If the left hand side is invariant with respect to a timelike
Killing field, and if $\l$ is also invariant, then so must be $u^b$.
Evidently the higher order terms in the equations of motion 
induce time dependence into the solution. A similar phenomenon
can be seen upon expanding the simpler Einstein-neutral dust system
about the flat space solution with constant dust 4-velocity and 
vanishing density.
As in our case, the dust density perturbation 
can set up any static metric perturbation, and the linearized 
geodesic equation for the dust yields a time-dependent dust
velocity perturbation.

\section{Matter couplings}
\label{matter}

We have so far considered only the terms in the
action involving the metric and the aether field and 
up to two derivatives. 
Suppose a matter term $S_{mat}[g_{ab}, u^a, \psi]$
is added to the action $S_{min}[g_{ab},u^a,\l]$ 
of the minimal theory (\ref{Smin}), where $\psi$ stands for
a generic matter field. The variation of $S_{mat}$ with respect
to the metric produces an additional contribution to the
stress-energy tensor, and the variation with respect to 
$u^a$ produces an additional term in the current
on the right hand side of (\ref{ueq}). The resulting 
field equation takes the form
\beq
\nab^a F_{ab}=-\frac{1}{2b_1}
\Bigl(\l u_b + \frac{1}{2}\frac{\d S_{mat}}{\d u^b}\Bigr).
\label{ueqmatter}
\eeq
The identity $\nab^a\nab^b F_{ab}=0$ then implies
\beq
u^a\nab_a\l=-\l \nab_a u^a 
-\frac{1}{2}\nab^a\frac{\d S_{mat}}{\d u^a},
\label{matterlambda}
\eeq
which shows that now even if $\l$ is initially zero it need not
remain zero. In the presence of such matter couplings the equivalence
to charged dust is lost. 

We now consider specific types of matter couplings, first of dimension
less than or equal to four, and next of dimension greater than four.

\subsection{Couplings of dimension $\le4$}
A complete classification of Lorentz violating,
gauge-invariant extensions of the 
SU(3)$\times$SU(2)$\times$U(1) minimal standard model
has been given by Colladay and Kosteleck\'y\cite{CK},
restricting attention to operators whose mass dimension
is less than or equal to four, so as to preserve 
power-counting renormalizability.  
This class of low energy effective actions includes
both $CPT$-even and $CPT$-odd terms, and involves 
various coupling tensors with ``generation"
indices allowing for mixing of fermions from different
generations. These coupling tensors are supposed to 
be Lorentz violating vacuum expectation values arising 
in a theory with a fundamental underlying Lorentz symmetry.

Here we consider the above class of Lorentz violating
terms, keeping only those couplings that can be constructed
with the aether field $u^a$. With this restriction the 
antisymmetric tensor couplings are excluded, 
which rules out Lorentz-violating Yukawa couplings 
and couplings of gauge field strengths to Higgs bilinears, 
and limits the form of
modifications of the gauge field kinetic terms. 
Invariance under time reversal
$u^a\rightarrow -u^a$ would be required if the physical
significance of the aether is only to define a preferred
frame and not a preferred local time orientation. 
If we accordingly further assume this symmetry,
all the $CPT$-odd terms are excluded, which rules out
terms with a vector coupled to fermion or Higgs currents,
gauge field Chern-Simons currents, and the U(1) potential.

The only possibilities remaining after all these restrictions
have been imposed are the modifications of the fermion,
gauge field, and Higgs kinetic terms:
\beq
\half i (c_L)_{IJ} u^au^b 
\bar{L}_I\g_a {D}_b L_J+ \mbox{h.c.}+ \dots,
\label{fermions}
\eeq
\beq
-\fourth c_B u^a u^b g^{mn}B_{am}B_{bn}+\dots,
\label{gaugefield}
\eeq
\beq
\half c_\Phi u^au^b(D_a\Phi)^\dagger D_b\Phi.
\label{higgs}
\eeq
The indices $I,J$ in (\ref{fermions}) are generational 
indices, and the coupling constants
$(c_L)_{IJ}$, $c_B$, and $c_\Phi$ are all dimensionless.
The ellipses in (\ref{fermions}) stand for similar 
terms for the other fermions,
while those in (\ref{gaugefield}) stand for similar terms
for the other gauge fields.
 
Such additional kinetic terms modify the
propagation speed of the various fields. For example, 
the propagation
speed for the Higgs field, with respect to the preferred frame, 
is $(1+c_\Phi)^{-1/2}$, which is less or more 
than the speed of light if $c_\Phi$ is positive or negative
respectively. The coupling
constants must therefore be small numbers for fields whose
propagator has been measured accurately.
It would be interesting to determine what limits can
be placed on these coefficients, particularly for
fields such as the Higgs or gluons whose propagators
are presumably not yet so well measured. 

\subsection{Couplings of dimension $>4$}

Once the restriction to terms of dimension 4 or less
is dropped, the possibilities for Lorentz violating
terms---like those for Lorentz invariant ones---are
endless. Here we would like to consider just two types,
which illustrate different possibilities that arise
in the presence of Lorentz symmetry breaking.

\subsubsection*{Modified kinetic terms}
If the coupling coefficient for a Lorentz 
violating kinetic term
like (\ref{fermions}-\ref{higgs})
is {\it field dependent} and polynomial, rather than
a constant, then the term is a dimension
$>4$ operator. In this case it is possible that
the coefficient was larger in the early universe
than it is today, due to the cosmological evolution
of the field(s) on which the coupling function depends. 
This provides an alternate approach to constructing
generally covariant, variable speed of light
cosmologies. Approaches using a vector\cite{moffatv} or 
a scalar\cite{moffats,vslc} to define the 
preferred frame have been the subject of some recent
papers. 

\subsubsection*{Modified dispersion}
\label{disp}
Next we consider a deviation from Lorentz invariance that 
becomes strong only at high wavevectors. In the early
universe, when the fields were highly excited at large
wavevectors, the gravitational effects of 
such a deviation could have been of paramount importance. 
The study of a model incorporating such effects 
is left to another paper\cite{nextpaper}. 
Here we indicate only an example of a
term in the Lagrangian 
that produces high frequency dispersion in
the propagation of a matter field, and we
display the the form of the resulting 
contribution to the energy-momentum tensor.

Consequences of non-Lorentz invariant high frequency
dispersion for the Hawking effect 
have previously been studied using 1+1 dimensional model
field theories 
in which higher spatial derivative terms are added to the 
action (for a review see \cite{river}),
and recently such models have been generalized to
field theory in the background of a 
3+1 dimensional Robertson-Walker spacetime in order to 
study the consequences for the spectrum of primordial density 
fluctuations in inflationary cosmology\cite{Martin}. 
These models can be extended to an arbitrary 3+1 dimensional
setting, preserving general covariance as well as spatial rotation
symmetry in the local preferred frame. As an example consider
the Lagrangian
\beq
{\cal L}_\phi=\frac{1}{2}\Bigl(\nabla^a\phi \nabla_a\phi
+k_0^{-2}(D^2\phi)^2\Bigr).
\label{Lmod}
\eeq
Here $k_0$ is a constant with the dimensions 
of inverse length which sets the scale for deviations from 
Lorentz invariance, and $D^2$ is the covariant spatial Laplacian,
i.e.,
\beq
D^2\phi=-D^aD_a\phi
=-q^{ac}\nabla_a(q_c{}^b\nabla_b \phi), 
\eeq
where $D_a$ is the spatial covariant derivative 
operator\cite{wald} and $q_{ab}$ is the spatial
metric (\ref{q}). 

The $u^a$-dependence of the Lagrangian
(\ref{Lmod}) produces a ``matter" term in
the aether field equation (\ref{ueqmatter}).
The energy-momentum tensor for this Lagrangian is
\begin{eqnarray}
&&T_{ab}=
\nabla_a\phi\nabla_b\phi-
{\cal L}_\phi\, g_{ab}\nonumber\\
&&\!\!\!- k_0^{-2}\Bigl[
2 D^2\phi\, u^m u_{(a}\nab_{|m|}D_{b)}\phi
+2 \nab_m(D^2\phi\, q_{(a}{}^m)\nab_{b)}\phi\nonumber\\
&&~~~~~~~~-\nab^m(q_{ab}D^2\phi\, D_m\phi)
\Bigr].
\label{tdisp}
\end{eqnarray}
In Ref. \cite{nextpaper} we evaluate the expectation value
of this energy-momentum tensor 
in a thermal state in flat spacetime, 
which allows us to determine
the modification of the equation of state produced by
the fourth derivative term. This equation of state is then
be used to study how the cosmological evolution is affected
by the high frequency dispersion. 

\section{Discussion}
\label{discussion}
We have made an initial attempt to study the possible consequences of 
incorporating a preferred frame---the aether---into a generally 
covariant theory. With the action adopted in this paper
the aether vector generically develops gradient singularities
even when the metric is perfectly regular. We take this as a
sign that the theory is unphysical as an effective theory
(although if the aether sector is ignored the theory can be 
made to agree with observations with an appropriate
choice of initial conditions, i.e. by setting $F_{ab}$ to zero.)
The primary open questions 
 are ($i$) what determines the initial values
of the aether field and the Lagrange multiplier field, and ($ii$)
are the gradient singularities, which appear to be generic in the evolution
of the aether, eliminated by including a symmetrized derivative term
$(\nab_{(a} u_{b)})(\nab^{(a} u^{b)})$ in the action along with
the antisymmetrized derivative term used in this paper? 
It is plausible that adding the symmetrized derivative term will
have a significant effect, since with it the action
is sensitive to the existence of any large gradients.

\section*{Acknowledgements}
We are grateful to V.A. Kosteleck\'y, M. Luty, and R.P. Woodard
for helpful discussions.
This work was supported in part by the National Science Foundation
under grant No. PHY98-00967.

\end{document}